\documentclass[aip,apl,groupedaddress,reprint]{revtex4-1} 
\usepackage{graphicx}
\usepackage{amssymb,amsmath,amsthm}
\usepackage{float}
\usepackage{tabularx}
\usepackage{array}
\usepackage{wrapfig}
\usepackage{color}
\usepackage{transparent}
\usepackage{cases}
\usepackage{ulem}
\usepackage{xcolor}
\usepackage{blindtext}
\definecolor{landanimal}{rgb}{.545,.1,.1}
\colorlet{ocean}{blue!60!black}

\newcommand{\transp}[1]{\ensuremath{#1^{\scriptscriptstyle \dag}}}
\usepackage{natbib}
\usepackage{bibentry}

\begin{document}

%%%%%%%%%%%%%%%%%%%%%%%%%%%%%%%%%%%%%%%%%%%%%%%%%%%%%%%%%%%%%%%%%%%%%%%%%%%%%%%%%%%%%%%%%%%

\title{Source and defect localization in thin elastic plates with arbitrary geometry}

%%%%%%%%%%%%%%%%%%%%%%%%%%%%%%%%%%%%%%%%%%%%%%%%%%%%%%%%%%%%%%%%%%%%%%%%%%%%%%%%%%%%%%%%%%%

\author{Martin Lott, Antonio S. Gliozzi and Federico Bosia}
\email[]{martin.lott@polito.it}
\affiliation{Department of Applied Science and Technology (DISAT), Politecnico di Torino, Torino, Italy}

\date{\today}

%%%%%%%%%%%%%%%%%%%%%%%%%%%%%%%%%%%%%%%%%%%%%%%%%%%%%%%%%%%%%%%%%%%%%%%%%%%%%%%%%%%%%%%%%%%
%%%%%%%%%%%%%%%%%%%%%%%%%%%%%%%%%%%%%%%%%%%%%%%%%%%%%%%%%%%%%%%%%%%%%%%%%%%%%%%%%%%%%%%%%%%

\begin{abstract}

In this paper, we experimentally demonstrate how discrete resonances can be used to image acoustic sources and mechanical changes in thin plates with different boundary shapes. The proposed methods uses coupled numerical with experimental data and requires only knowledge of the sample geometry. The free modes of the plates are not orthogonal from the receivers' point of view, which induces an artificial coupling in the post-processing of the experimental signals. However, we show that this effect can be corrected using numerical simulations and a mathematical transformation of the antenna geometry. After this correction, imaging of active sources is performed using coherent summation of the elastic field over the natural frequencies of the plates, leading to an unique possible localization of the sources. Imaging mechanical changes in the two plates, however, is addressed using incoherent summation over the modes, leading to symmetry problems for these nearby cavities. This work experimentally illustrates the spatial resolution, perspectives and limitations in the use of eigenmodes to produce images in complex elastic systems.

\end{abstract}

\maketitle

\section{Introduction}

The imaging of an object using elastic waves relies on the ability to discern any point in space by recording and processing acoustic data. For example, if the bulk density $\rho$ and sound celerity $c$ of a fluid are known, wave diffraction theory can predict in any point and at any time the acoustic wave field inside the medium from measurements at the boundaries \citep{baggeroer1988matched}. This theory has led to the development of advanced array imaging techniques used in different fields of research, from wave physics applications \citep{prada1996decomposition, aubry2009detection, aubry2014recurrent,larose2010locating} and medical imaging \citep{robert2008green, fasi2015algorithm} to seismology \citep{colombi2014temporal, touma2021distortion, poli2017analysis}. 

Using matrices to represent and analyze spatiotemporal acoustic data has the advantage to abstract complex operations like space filtering, beam-forming and wave polarization analysis \citep{poggi2010estimating, seydoux2017pre}. In addition, this type of formalism bridges the gap between an analytical formulation and experimental data manipulation, especially with the use of synchronized source/sensor arrays, where the backpropagation operators in space and time can be treated as simple matrix multiplications \citep{lambert2020distortion,baggeroer1988matched, collins1991focalization,fink2000time}. Usually, diffraction theory for imaging is adopted when the elastic wavelengths are much smaller than the domain to image. This limit helps to remove potential unwanted wave reflections and conversions from boundaries, which strongly affect the quality of an image \citep{aubry2009detection}. 

In the low frequency limit, when the wavelengths are of the same order of magnitude as the domain to be probed, the waves do not propagate and form the so-called standing wave, with discrete resonance effects over frequency. Imaging a medium using its resonant modes is also a wide field of research in itself. Modal Analysis (MA) techniques (also known as System Identification) are used in civil \citep{astorga2019recovery}, mechanical \citep{senjanovic2016analytical} and aerospace \citep{agneni1996damage} engineering. They are deployed on systems and structures that presents complex geometries and boundary conditions, meaning situations where classical diffraction theory cannot simply applied \citep{roux2014structural}. One of the main advantages of MA is that the resonant modes are extended in space. This means that the entire structure is affected by the vibration, with node and antinode locations that are frequency dependent. From an energetic point of view, the lowest frequency modes of a structure gather most of the mechanical energy that the system receives. This phenomenon is also used to discern the presence of potential defects in a structure, which is very sensitive to this level of vibrational energy\citep{cawley1979location, farrar1998comparative, humar2006performance, eiras2021damage, van2007multi, lott2018three}. However, previous research has shown that the curvature of modal shapes (i.e., the spatial derivative of the modal displacement), and not the resonant frequencies, is the most effective tool for imaging\citep{eiras2021damage, van2007multi, gliozzi2006modelling}, even for a simple one-dimensional structure\citep{van2007multi, roux2014structural}. This is a strong limitation for the use of modal methods, which consequently require significant instrumentation and mainly surface imaging capabilities.\\

In the present paper, we propose an experimental realization of modal elastic imaging in an arbitrary elastic system, which draws on both modal analysis and wave diffraction paradigms. The methodology relies on the simultaneous manipulation of multiple discrete modes (eigenmodes), which are presents in a highly reverberant system. The natural frequencies and modal shapes are determined numerically, from a data-driven procedure only requiring knowledge of the sample geometry. Then, appropriately combining vibration data and numerical modeling, two imaging problems are addressed: the imaging of an active source and the imaging of a small defect in a reverberant environment. To assess the role of symmetries in the imaging procedures, the experimental setup include two different thin aluminum plates with different boundary shapes (rectangular and irregular pentagonal). 

The paper is organized as follows. First, we present the studied samples and the procedure for acquiring the vibration data. Second, we describe the numerical modeling and the details of the signal processing algorithm using synthetic data for illustration purposes. Third, the experimental source localization images are presented for both samples and compared to the synthetic data. Finally, results and perspectives for defect-like imaging with monitoring techniques using discrete resonant modes are discussed. 

\section{Materials and methods}

\subsection{Experimental set-up}
%The experiments consist in recording at 16 random locations in space, the temporal evolution of the wave field from 3 different and independent active sources. The main purpose of this first experiment is to image the source location, from the analysis of the highly reverberated time signal at the receivers locations. The data processing is based on a universal Green's function expansion method, which uses the only knowledge of the normal modal shape and the associated eigenfrequency to derive the spatiotemporal evolution of the wave field inside the plates. We show that from the recording array point of view, the eigenmodes of the plates are not orthogonal and we propose a remediation. In addition, we introduce a phase locking methods, capable of identifying the emission time of an active source, from the spatiotemporal analysis of the reverberated coda signal. 

\begin{figure}
    \includegraphics[scale=0.55]{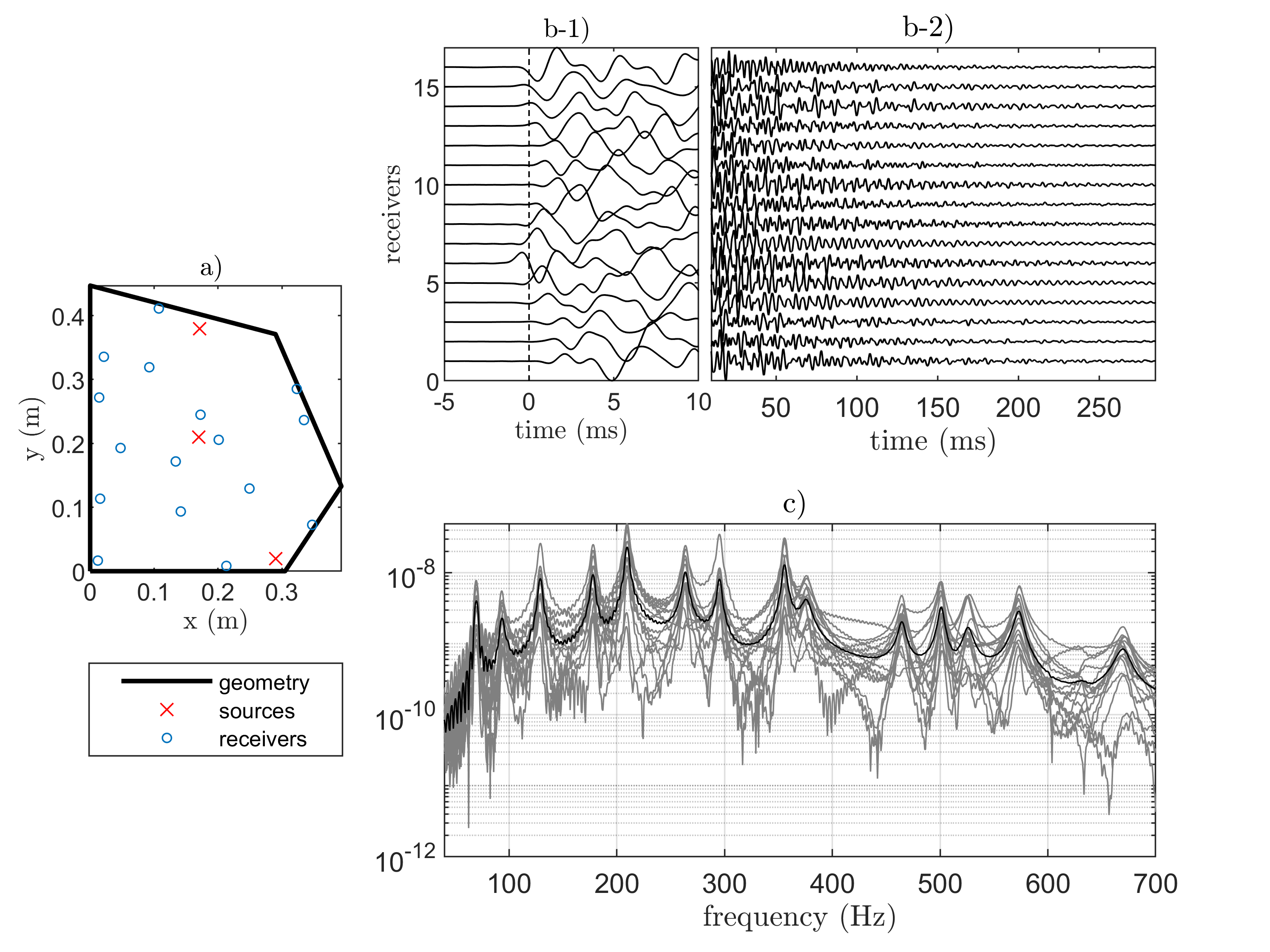}
    \caption{\label{fig:signals} Irregular plate geometry with its instrumentation. a) For each of the three sources, 16 signals are recorded at the blue location points with a Doppler laser vibrometer, b) with a 300 ms of a strongly reverberated signal (coda). c) In the frequency domain, the signals show clear and isolated resonance peaks. The average of the spectra over the 16 measurement points is represented by the black line.}
\end{figure}

The first sample used in this study is a thin aluminum plate of 3 mm thickness with irregular and non-parallel edges (Fig \ref{fig:signals}). The particular shape given to the plate guarantees the absence of degenerate modes (i.e. modes with the same resonance frequency but with different modal shapes). Since this study aims to exploit modal imaging concepts, we consider the response at the lowest frequency of the plate. In this frequency regime, the only admitted mode within the plate is the out-of-plane, antisymmetric, dispersive $A_0$ Lamb mode. For a 3 mm plate, we can determine the wave velocity of this mode from the formula :

\begin{equation}
C_p = \sqrt[4]{\omega^2\dfrac{D_0}{h\rho}} ;
\end{equation}

Where $D_0 = h^3E/(12(1-\nu^2))$ is the flexural rigidity, and $E$, $\nu$, $\rho$ and $h$ are the Young's modulus, Poisson's ratio, bulk density and thickness, respectively. The wavelength is of the order of $\lambda = 50$ cm at 100 Hz, which roughly corresponds to typical plate lateral dimensions (see fig. \ref{fig:signals}-a), meaning that the 10 lowest modes of this plate should be located around 100 Hz.\\

The experiment consist in recording at different random locations on the sample, the temporal evolution of the wave field generated by vaious independent active sources. The main purpose of this first experiment is to image the source location, from the analysis of the highly reverberated time signal at the receiver locations. The plate is thus instrumented with 3 piezoelectric disks (ABT-441-RC9, 4.2 kHz central frequency) with a diameter of 27 mm and a thickness of 0.5 mm, which act as the sources. The piezoelectric disks are independently driven with a 2 second chirp signal, with a frequency content ranging from 20 Hz to 900 Hz. This signal is generated by an arbitrary signal generator (Agilent 33500 B), and amplified with an high impedance output tension amplifier (FLC Electonics A400). After propagating in the plate, the vibration response is measured using a Doppler laser vibrometer (Polytech OFV-500 decoder and OFV-505 sensor-head) at 16 different plate locations. The recorded signal is then cross-correlated with the source signal, to provide a 0.5 s highly reverberated impulse response. For each sources, the obtained data-set can be represented by a matrix $\mathbf{K_{r}(t)}$, indexing time at sampling rate intervals and space $r=1, 2, .. 16$ from receiver locations in two dimensions. A schematic of the experiment is shown in Fig. \ref{fig:signals}-a. Two sources are located on the edges of the plate and one in the center. The receivers are randomly distributed over the plate. A typical example of the 16 recorded signals and their corresponding frequency spectra for a single source are shown in Fig. \ref{fig:signals} b-c. 

From Fig. \ref{fig:signals} panel b-1, one can see that as soon as the signal is emitted, reflected waves appear that mask any potential coherent propagation. From (fig. \ref{fig:signals} panel c), the Fourier transforms highlight separated peaks, corresponding to the first resonant modes of the plate. 

\subsection{Modal shape determination}

The proposed approach exploits knowledge of the modes of vibration. There is in the literature a large set of methods able to compute normal modes and associated frequencies for systems with simple geometries \cite{wang2004static, migliori1993resonant, mizusawa1986natural, poli2017analysis}. Usually, these methods are based on Hamilton's principle of least action, according to which a dynamic system tends to minimize its internal mechanical energy. This minimization principle leads to an eigenvalue problem, with the resonance frequencies as eigenvalues and the normal vibration mode shapes as eigenvector solution of the problem \cite{migliori1993resonant}. For the considered plate geometry with Neumann boundary condition (free edges), analytical models run the risk of non-exhaustively predicting the eigen-modes \citep{senjanovic2016analytical}. One solution is to use finite element modeling (FEM) and eigenvalue analysis to obtain the eigenmodes and corresponding eigenfrequencies. The input mechanical parameters (Young's modulus and Poisson's ratio) can then be adjusted to find the best match between the measured and numerically predicted resonance frequencies. This procedure is also know under the name of Resonant Ultrasound Spectroscopy (RUS)\cite{migliori1993resonant, zadler2004resonant, remillieux2015resonant}.

The numerically calculated and experimentally derived normal modes should coincide for the source localization procedure to be successful. A way to guarantee this similarity is to perform a systematic spatial projection of numerical modes on the measured ones. This can additionally improve the RUS methodology by simultaneously optimizing the agreement between experimental and numerical data for both the resonant frequencies and modal shapes. To do this, we consider the array response matrix of the three sources $s$ in the frequency domain $\mathbf{\Tilde{K}^{(s)}(\omega)}$, projected over the receivers array on the modal shape $\phi_{r n}$. 

\begin{equation} \label{eq:projection}
    B_n(\omega) = \Sigma_s\left|\Sigma_r\Tilde{K}^{(s)}_r(\omega) \phi_{r n}\right|
\end{equation}

\begin{figure}
    \includegraphics[scale=0.5]{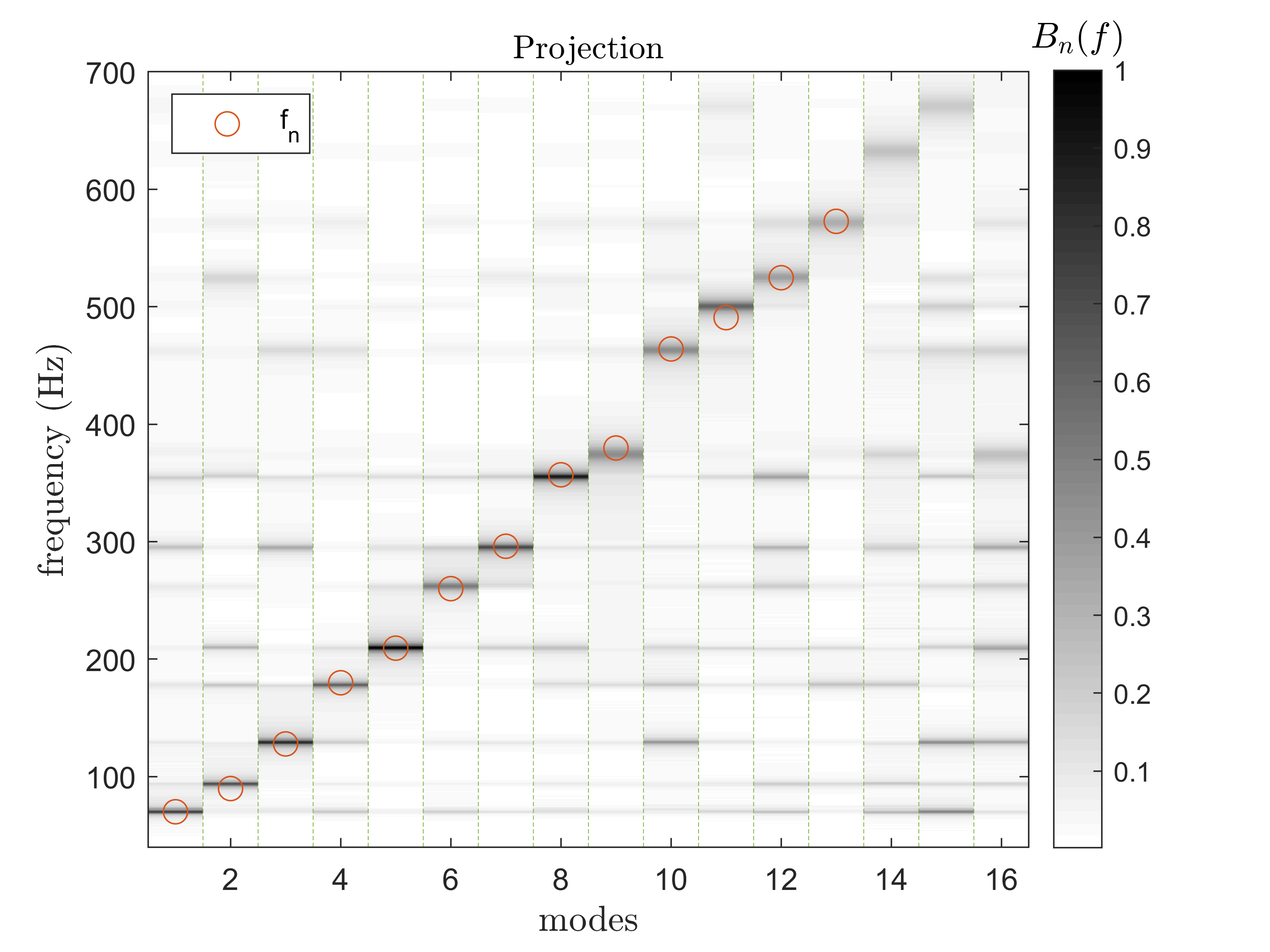}
    \caption{\label{fig:inversion}Frequency response of the normalized modal projection. The red circle represents the eigenfrequency values computed with FEM. The gray-scale represents the continuous response over frequency for each mode, projected (eq.\ref{eq:projection}) on the FEM eigenmodes.}
\end{figure}

The starting guess for the mechanical parameters values are $E=69$ $GPa$ for the Young's modulus, $\nu=0.33$ for the Poisson ration, and $\rho=2700$ $kg.m^{-3}$ for the density. Then, for different Young's modulus and Poisson's ratio values around the starting point, we compute the eigenfrequencies and eigenmodes of the system, and consider the minimization function :

\begin{equation}
    g = \Sigma_n \left|\omega^d_n- \omega_n\right| / \omega_n
\end{equation}

with $\omega^d_n$ the angular frequency with the maximum amplitude response of the array projected on the eigenmode $\phi_{r n}$, and $\omega_n$ the numerical eigenfrequency of this same mode computed from FEM. In this manner, we include the spatial response of the modes in the inversion procedure. Results of the projection defined in Eq.\ref{eq:projection} are depicted on in Fig. \ref{fig:inversion} for optimized mechanical parameters. This figure illustrates the post-processing mode separation by means of the projection between the experimental data and the numerical eigenmodes. With 16 receivers, the first 13 modes present a maximal projection value $B_n(f=\omega/2\pi)$ at the corresponding eigenfrequency $f_n=\omega_n/2\pi$. Final values are $E=69.6$ $GPa$ for the Young's modulus and $\nu = 0.345$ for the Poisson ratio. Note that an error in the determination of the geometry of the sample could be translated in an uncertainty in the estimated mechanical parameters. 

\begin{figure}
\includegraphics[scale=0.65]{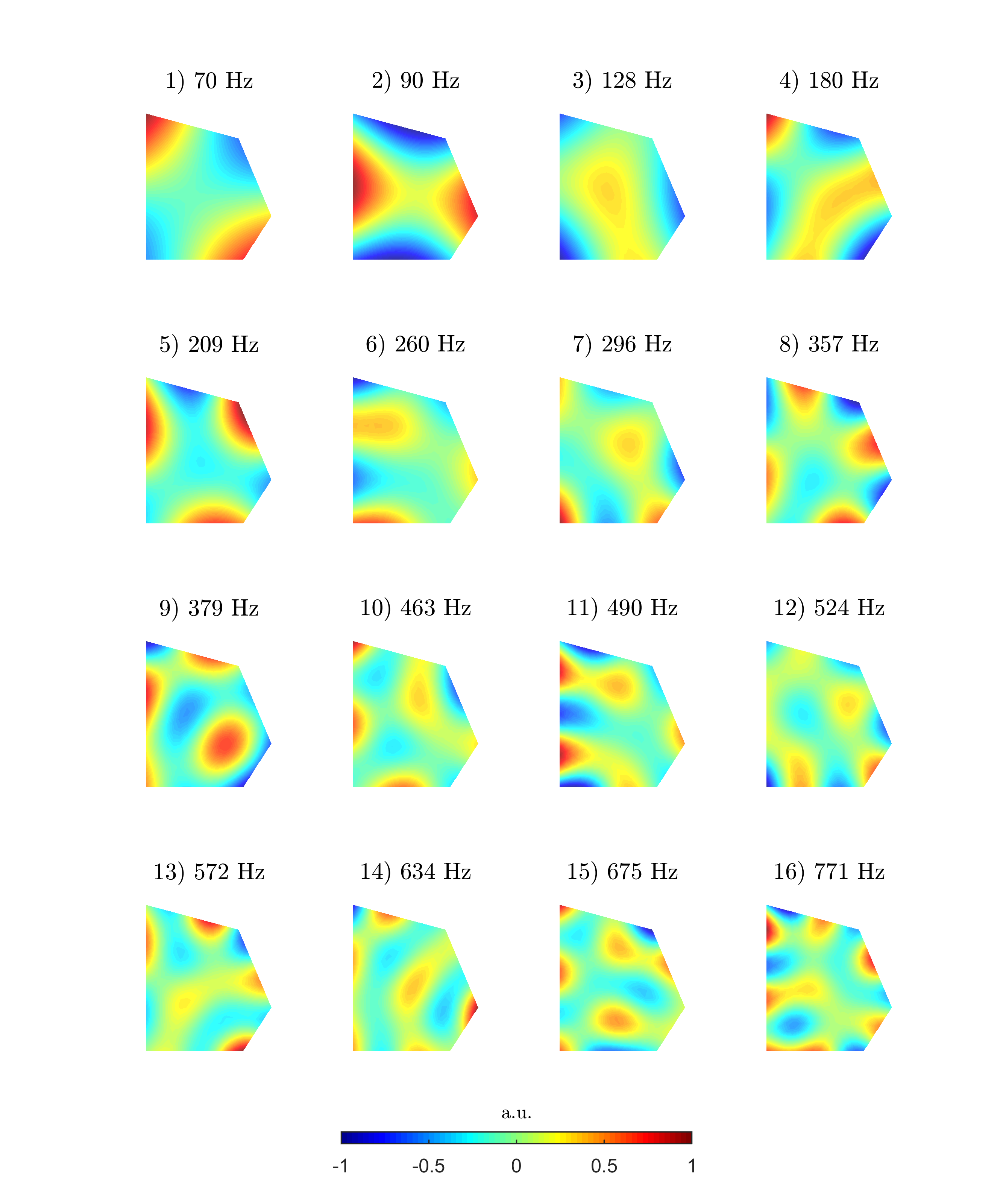}
\caption{\label{fig:modes1}The 16 first modal shapes of the aluminum plate obtained with FEM.}
\end{figure}

The normalized motion profile of the first 16 eigenmodes are shown in Fig \ref{fig:modes1}. The sharp corners used for the design of the plate create a clear separations between modes. From the modal fields, the maximum displacement zones appear to be located at the edges of the plate, whilst the nodes of vibrations are principally located in the center of the sample. 

\subsection{Modal Green's function expansion and source localization principle} 

The the spatiotemporal evolution of the wave field inside the plate is derived using a universal Green’s function expansion method, which only requires knowledge of the normal modal shapes and the associated eigenfrequencies. We start from the expression of the Green's function in the time domain\citep{weaver1994weak,gallot2011coherent, catheline2011coherent}:

\begin{equation}
    G_{ij}(t) = \Sigma_n \phi_{i n} \transp{\phi_{n j}} \cos(\omega_n t) 
\end{equation}

Where $i,j$ denote the location of two points in the continuous sample domain $\Omega$, $n$ designates the modal index, $\omega_n$ the angular eigenfrequency and $\phi_{in}$ the n-th eigenmode amplitude at the point $i$. The time Fourier transform of the Green's function, computed from $t=0$ to infinity can be expressed as follows :

\begin{equation} \label{eq:FourierGreen}
    \tilde{G_{ij}}(\omega) = \Sigma_n \phi_{i n} \transp{\phi_{n j}} \times \delta(\omega-\omega_n)
\end{equation}

Where $\delta(\omega-\omega_n)$ is the continuous delta function. From this expression, we can express the broadband Green's function, with a continuous integration over frequency:

\begin{equation}\label{eq:greenexpans}  
    G_{ij} = \int_0^\infty \tilde{G_{ij}}(\omega) \text{d}\omega = \phi_{i n} \transp{\phi_{n j}} 
\end{equation}

Where we have used Einstein’s convention for summation over repeated indices. With current notations, one should verify:

\begin{align}\label{eq:ortho}
    \phi_{in} \transp{\phi_{n j}} &= \delta_{ij} \\
    \transp{\phi_{n i}}\phi_{i n'} &= \delta_{nn'}
\end{align}

Where $\delta_{ij}$ is the discrete delta function. The first equation indicates that the finite sample response to a source is unique and only depends on its geometrical location. The second equation illustrates the spatial orthogonality of the modal shape functions.

\begin{figure}
    \includegraphics[scale=0.65]{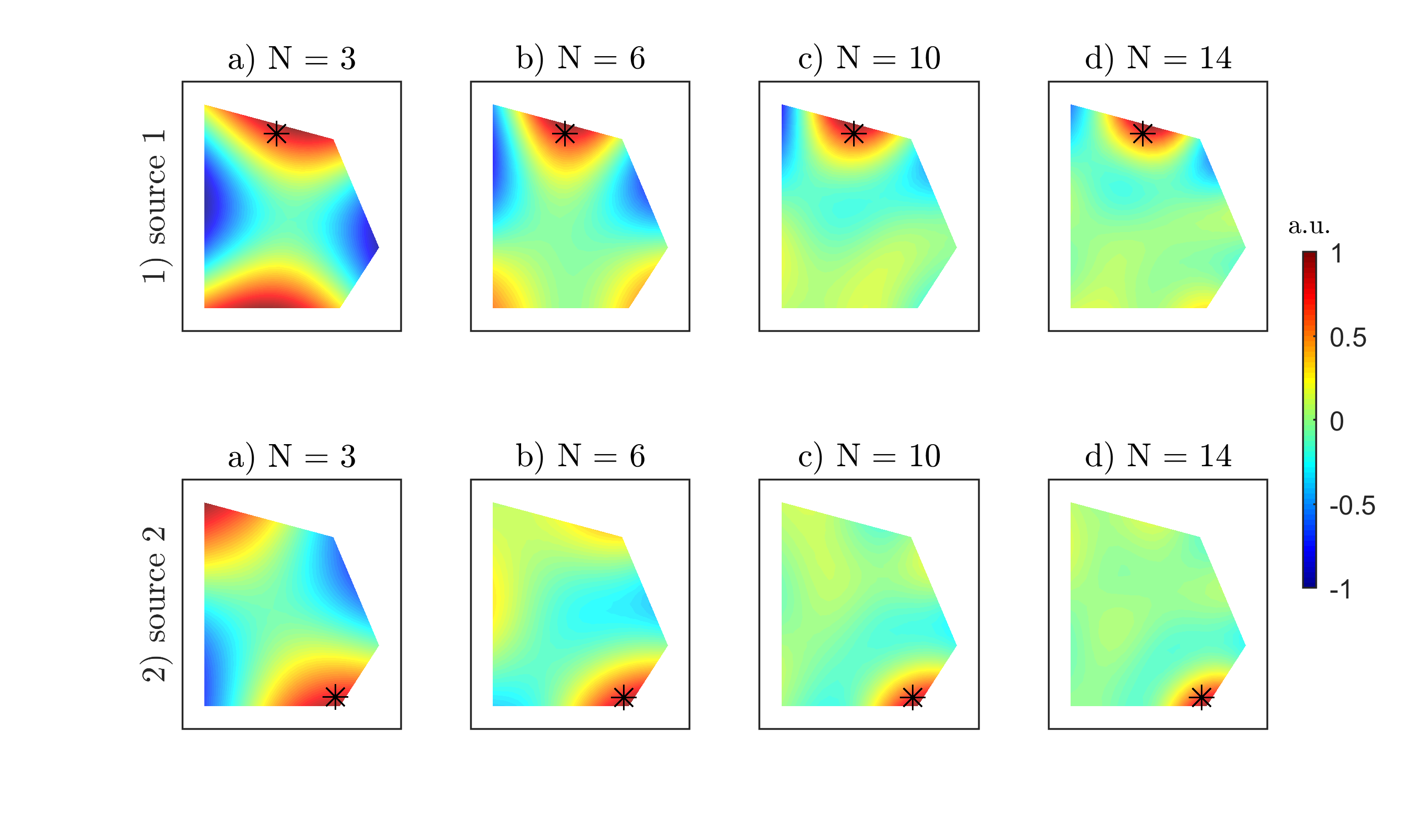}
    \caption{\label{fig:sourcevsmodes}Synthetic source reconstruction for two different location (1-2) and using 3, 6, 10 and 14 modes respectively.}
\end{figure}

Using these concepts, and the modal responses shown in Fig. \ref{fig:modes1}, we evaluate the effect of the number of modes on the source reconstruction. From equation \ref{eq:greenexpans}, the source image is obtained by multiplying the modal amplitude vector at the source location $\phi_{in}$, with the modal response matrix at each point location of the sample $\transp{\phi_{n j}}$. Results are shown in Fig. \ref{fig:sourcevsmodes}. As expected, increasing the number of modes increases the resolution of the source reconstruction. In addition, the location of the sources also affects the sharpness of the focus.

The central problem of imaging the source lies in the fact that from an experimental point of view, equation \ref{eq:ortho} is not satisfied when using a small subset of measurement points in the continuous $\Omega$ domain. However, it is still possible to construct an image of the source using the composition of Green's function within a set of virtual sources and receivers points $r$:

\begin{eqnarray}
     G_{ij} &=& G_{ir} G_{rj} \\ \label{eq:reconstruction1}
     &=& \phi_{in}\transp{\phi_{n r}}\phi_{rn'}\transp{\phi_{n'j}} \\
     &=& \phi_{in} F_{nn'}\transp{\phi_{n' j}}
\end{eqnarray}

The matrix $\mathbf{F}$ symbolize the ability of any subsequent set of points $r$ inside the $\Omega$ domain to independently identify each mode. If the implicit summation over $r$ is performed on the full domain $\Omega$, the Green's function composition is indeed equivalent to equation \ref{eq:greenexpans}, and we must verify that :

\begin{equation} \label{eq:visibilitymatrix}
    F_{nn'} = \delta_{nn'}
\end{equation}

The projection error due to the limited number of receivers can be corrected. Indeed, with a finite number of modes and receivers, we can determine a new ${\psi_m}$ basis, constructed as linear combination of the modal shapes ${\phi_n}$. This new basis must be orthonormal from the point of view of the receivers. The construction operation corresponds to the qr-decomposition, which establishes that any rectangular matrix can be decomposed into the product of an orthogonal matrix $\mathbf{Q}$ and an upper triangular matrix $\mathbf{R}$:

\begin{equation}
    \transp{\phi_{n r}} = Q_{n m} R_{m r}
\end{equation}

Where \textbf{R} matches the receivers to the new basis $\{\psi_m\}$, constructed as linear combination of the modal basis $\{\phi_n\}$. \textbf{Q} matches the new basis $\{\psi_m\}$ to the modal basis $\{\phi_n\}$. If $\mathbf{R}$ is invertible, the qr-decomposition is unique. 

The expression of the Green's function $G_{ir}$ in the new frame basis become :

\begin{equation}
    G'_{i\{\psi_m\}} = G^{}_{ir}R^{-1}_{r m}
\end{equation}

Using the correspondence between the $\{\phi_n\}$ and $\{\psi_m\}$ basis, and back-project the wave field into each pixel of the image, we can write a potential Green function $G'_{ij}$ as

\begin{equation}
    G'_{ij} = G^{}_{ir}R^{-1}_{rm}\transp{Q_{m n'}}\phi^{}_{n'j} \label{eq:reconstruction2}
\end{equation}

We can prove that this expression is the Green's function reconstruction, from the source $i$ to each pixel $j$, using only $r$ points in the domain $\Omega$. It is valid if the inverse of the matrix $\textbf{R}$ exists. This means that $\textbf{R}$ needs to be (but this is not a sufficient condition) a square matrix (i.e. same number of receivers and modes).

\begin{equation}
    G'_{ij} = G_{ir}R^{-1}_{rm}\transp{Q_{mn'}}\transp{\phi_{n'j}} = \phi_{in}\transp{\phi_{n j}} = G_{ij}
\end{equation}

\begin{figure}
    \includegraphics[scale=0.55]{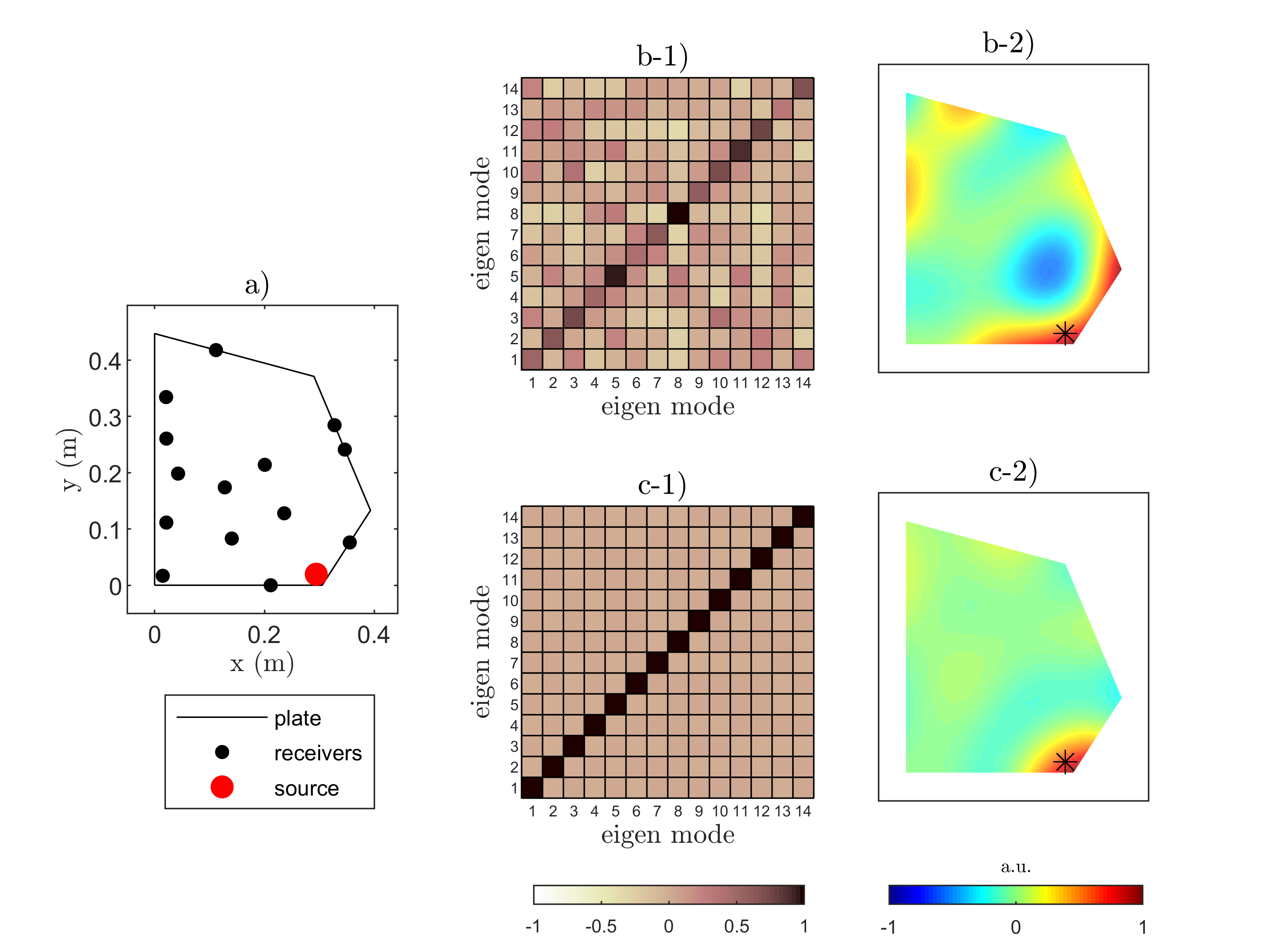}
    \caption{\label{fig:reconstuctionppe}Principle of source reconstruction with synthetic data. a) Specimen geometry with virtual receivers and source location. b) Using direct Green's function composition leads to a distortion of the matrix $F_{nn'}$ (Eq. \ref{eq:visibilitymatrix}) (b-1) and errors in the reconstructed signal (b-2). c) Using the array correction, each virtual mode is independent (c-1) and returns an exact image of the source, for the considered modes combination (c-2).}
\end{figure}

It is possible to validate the approach with synthetic signals. Here, we limit the example to a single source location, detected with 14 receivers, randomly located over the sample, as shown on fig. \ref{fig:reconstuctionppe}-a. We first compute the direct, non-corrected, Green's function composition that we applied to the antenna made of the 14 receivers, following eq. \ref{eq:reconstruction1}. Results are shown in fig.\ref{fig:reconstuctionppe} b-1 for the matrix $F_{nn'}$ and b-2 for the reconstructed field. The presence of non-zeros values on the off-diagonal elements of the $F_{nn'}$ matrix in fig.\ref{fig:reconstuctionppe} b-1 quantifies the cross-talk between the normal modes due to the limited number of receivers. This leads to spatial errors in the source imaging process. On the other hand, if we apply the qr-decomposition and backproject it using modified basis (Eq. \ref{eq:reconstruction2}), cross-talk between modes is eliminated and source reconstruction is considerably improved (Fig. \ref{fig:reconstuctionppe} c-1 and c-2). Here, signal reconstruction is performed using only a number of receivers equal to the number of modes ($N=14$).

\section{Results}

\subsection{Phase locking}

Application of the proposed method to real data requires incorporating the effect of time into the modeling. At time $t=0$, when the broadband pulse is transmitted, each frequency making up the signal, as a starting point, acquires the phase of the modal shapes at the source location. For stationary modes, this original phase is equal to 0 or $\pi$. If this phase information is lost, i.e. we no longuer know when the pulse was emitted, the resulting sum in equation \ref{eq:greenexpans}, leads to an incoherent summation of the modes and to the loss of spatial resolution. A general method for the inversion of the time origin of an active source is presented in \citep{collins1991focalization}. We propose here a simpler solution to this problem, exploiting the fact that the space-time Fourier transform of the receiver array is real, according to Eq. \ref{eq:FourierGreen}.

$\mathbf{K_r(t)}$ is the matrix of receivers' data, which records the temporal coda (reverb) signal from a single source in the plate, transmitted at time $t=0s$ but recorded starting at an unknown time $t_0>0s$. For each resonant mode, we consider the delayed-time Fourier transform of the spatial projection, computed from time $t=t_0$ over a time window $T$:

\begin{eqnarray}
    && \tilde{F}_n(\omega, t_0-\tau) =\nonumber \\
    && \exp{\left(i\omega(t_0-\tau)\right)} \int_0^{T} K_r(t)\phi_{rn}\exp{\left(i\omega t\right)}\text{dt}
\end{eqnarray}

Since $\tilde{F}_n(\omega, t_0-\tau)$ should be real for all $n$, we look for the minimum of the function :

\begin{equation}
    h_{t_0}(\tau) = \dfrac{1}{N} \Sigma_n \dfrac{\left| \Im\left\{\tilde{F}_n(\omega_n, t_0-\tau)\right\}\right|}{\left|\tilde{F}_n(\omega_n, t_0-\tau)\right|}
\end{equation}

\begin{figure}
    \includegraphics[scale=0.6]{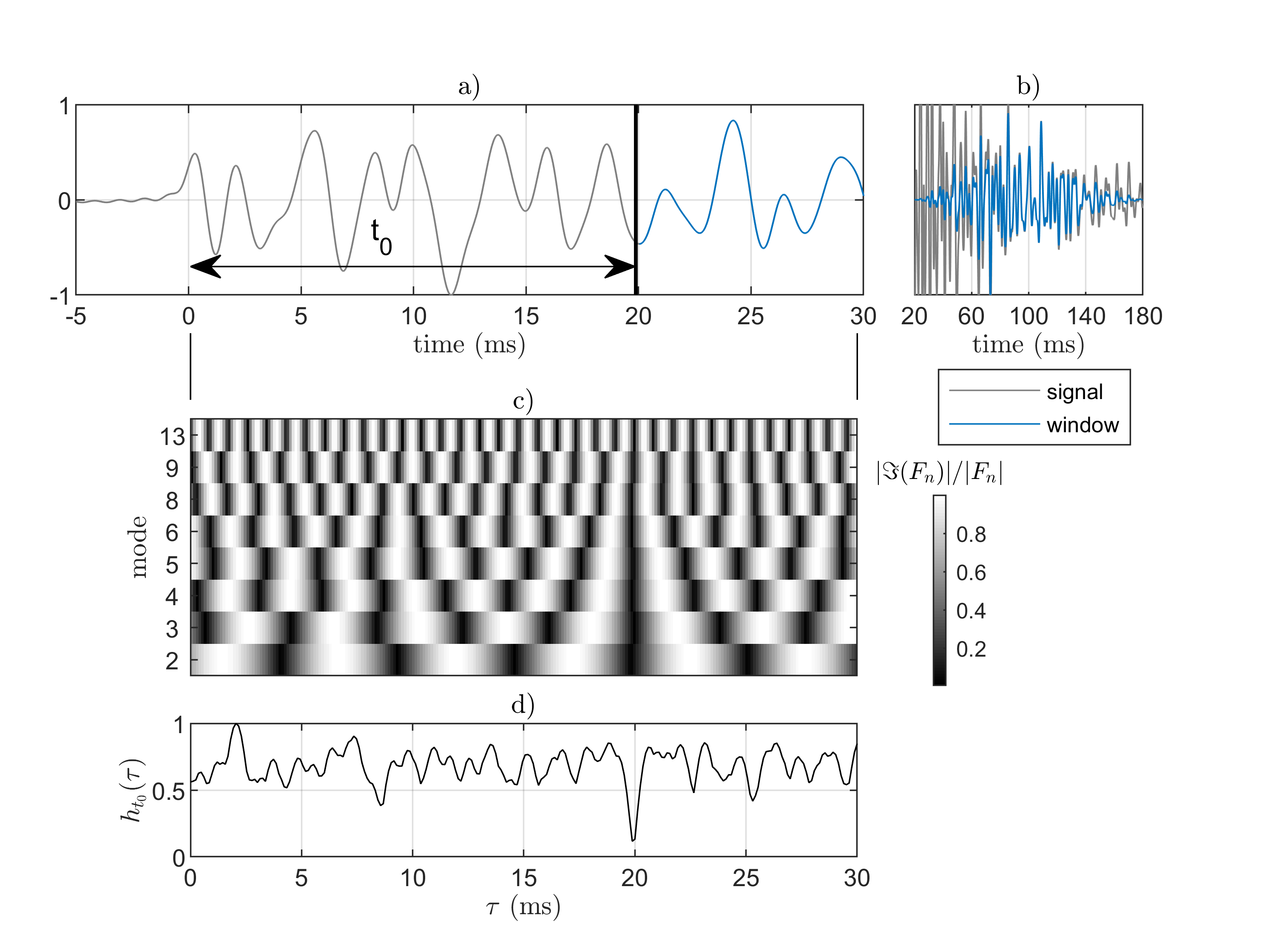}
    \caption{ \label{fig:lockingphase}Phase locking method. a) Time signal emitted from source 2 and recorded at location 5. The considered time window for the analysis starts at time $t_0=20$ $ms$, highlighted with the black vertical line. b) Hanning time window considered for the analysis (blue signal) of length $T=160$ $ms$ in the late coda part of the total transmitted signal (gray signal). c) Imaginary part contribution of the Fourier transform for 8 modes with the highest signal to noise ratio. d) At $\tau=t_0=20$ $ms$, all the modes are in sync, which corresponds to a minimum for the function $h_{t_0}(\tau)$}
\end{figure}

Where $\Im$ indicates the imaginary part operator. This procedure is graphically illustrated in Fig. \ref{fig:lockingphase}. In panel a), the signal emitted from source 2 to receiver 5 is shown. From this signal, we apply a Hanning time window in the late part of the coda, starting at $t_0=20$ $ms$, shown in panel b). For eight of the modes with the best signal to noise ratio, the function $\tilde{F}_n(\omega, t_0-\tau)$ is computed and depicted in panel c). For $\tau=20$ $ms$, all modes are synchronized, which corresponds to the delay $\tau=t_0$. The sum over the modes $h_{t_0}(\tau)$ is presented in sub panel c). Thus, even if $t_0$ (i.e. the temporal origin of the signal) is unknown, it is possible to trace back the origin of the signal using the spatiotemporal analysis of the coda. This is used as a "phase locking" method on the data.

\subsection{Experimental source localization}

\begin{figure}
    \includegraphics[scale=0.6]{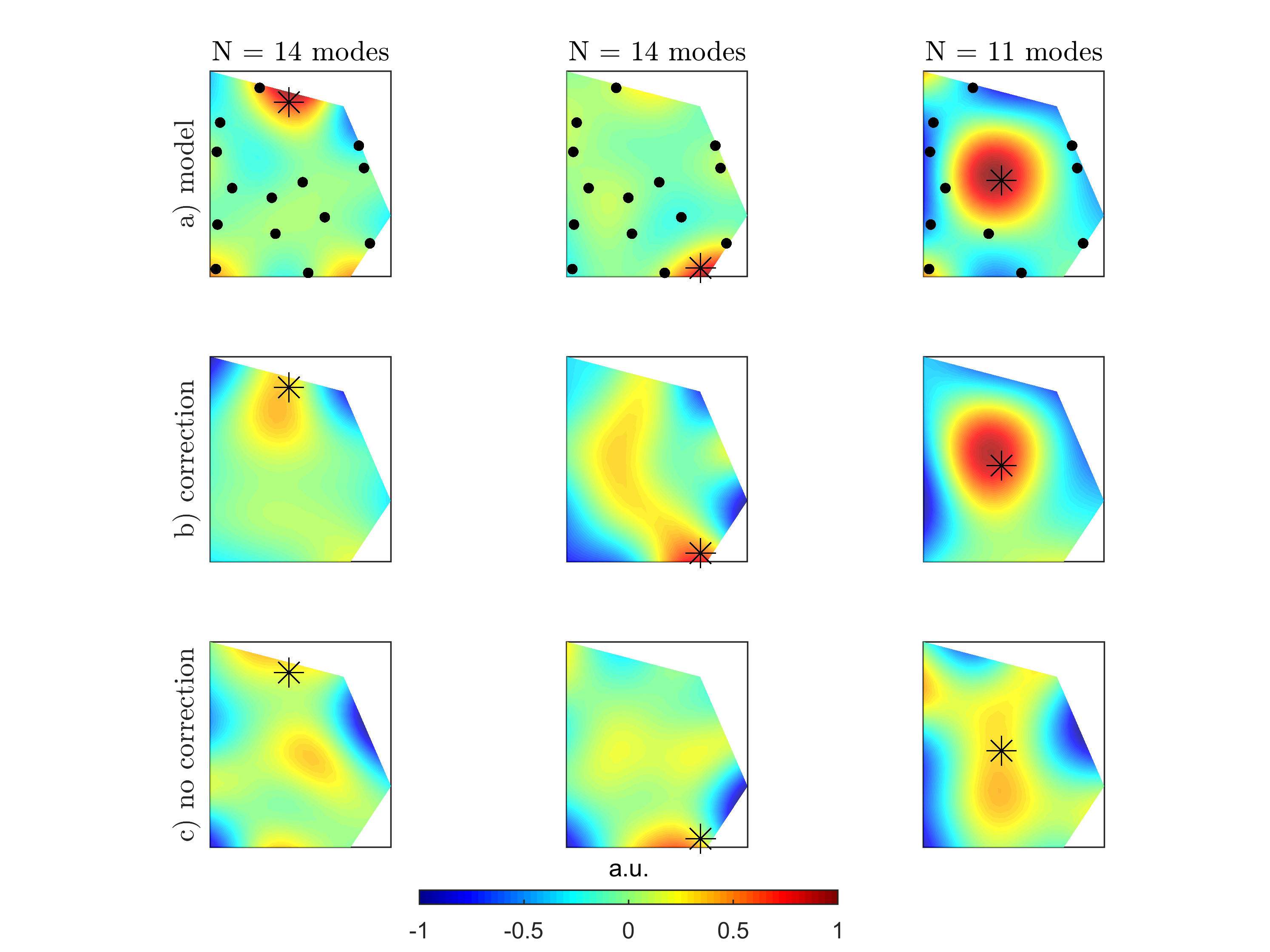}
    \caption{\label{fig:imagessources} Images of the reconstruction. panels a) show the expected image with the used receivers as black points and the source as a black star. b) the main resulting image after corrections. Panels c) show the results without correction, for comparison purpose. The images are normalized by the maximum of their absolute value.}
\end{figure}

In order to experimentally determine the source location in the considered plate, the real part of the Fourier transform after the phase correction is summed over the eigenfrequencies and the resulting vector is backpropagated according to Eq. \ref{eq:reconstruction2}. The theoretical reconstruction model for three examples of source location on the considered plate is displayed Fig. \ref{fig:imagessources}-a, experimental results are shown in Fig. \ref{fig:imagessources}-b, and the reconstruction without correction in Fig. \ref{fig:imagessources}-c. The images in Fig. \ref{fig:imagessources}-b show that the reconstruction is effective in determining the location close to the expected theoretical image (Fig.\ref{fig:imagessources}-a), although the resolution is in some cases limited. In Fig.\ref{fig:imagessources}, we only use a reduced number of modes with the highest signal to noise ratio in the reconstruction and adapt the array size to keep the possibility to compute the inverse matrix $\mathbf{R}^{-1}$ in equation \ref{eq:reconstruction2}. The reduction of the number of useful modes is due to the imperfection of the source, which does not provide a perfectly flat response in the considered frequency range. In addition, if the source is located on a vibration node of the plate, this mode can not be excited. These are limiting factors in the efficiency of the method. In the present work, we use the modal amplitude directly estimated in the late coda part of the signal, without further post-processing and we only consider modes with the highest signal to noise ratio. A more accurate approach would need to invert location and modal normalization at the same time.

\subsection{Application to regular geometry}

\begin{figure}
    \includegraphics[scale=0.6]{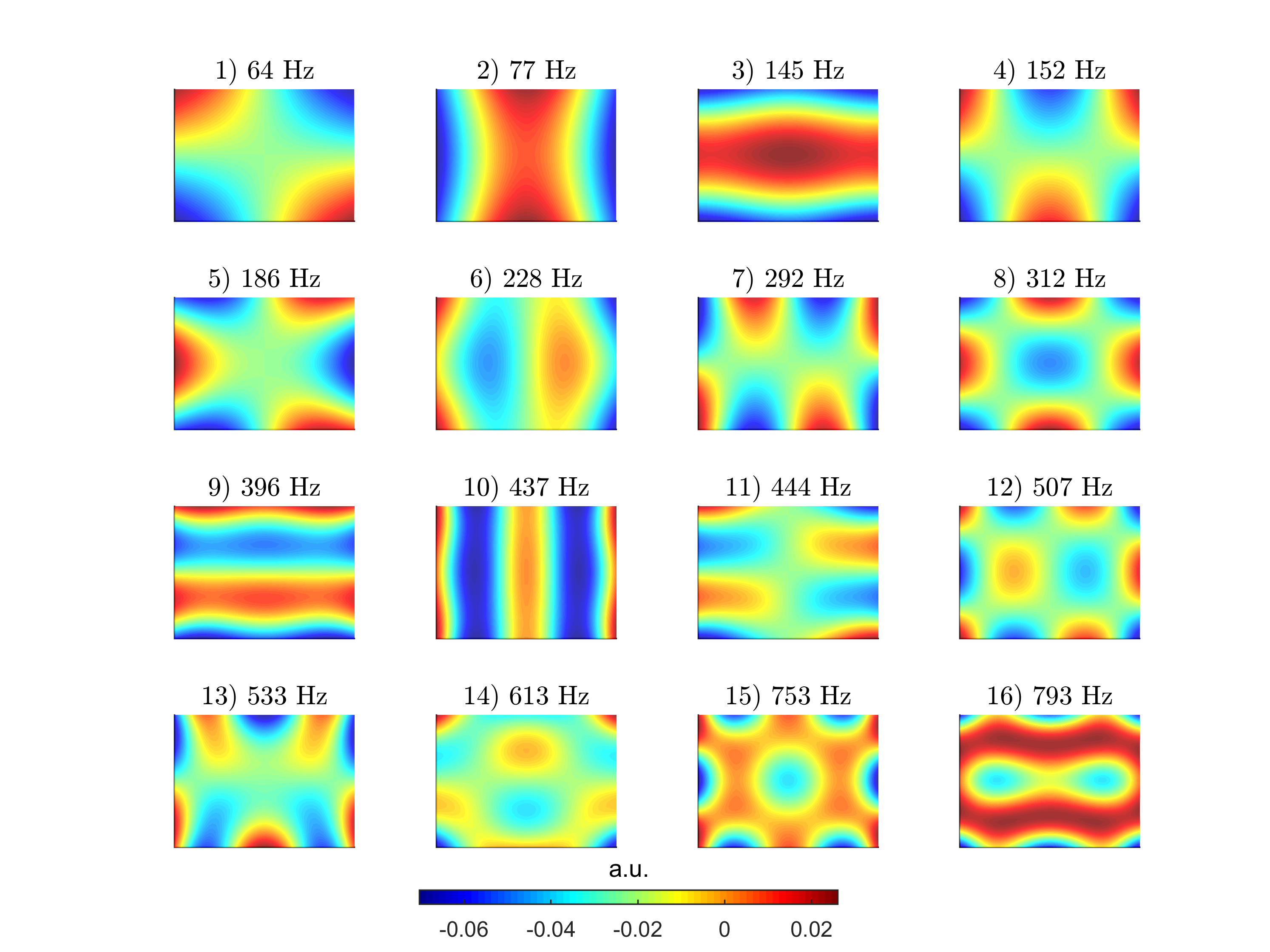}
    \caption{\label{fig:modesrectangular}The 16 first modal shape of the rectangular aluminum plate}
\end{figure}

To evaluate the influence of specimen geometry, we produce the same data-set as previously on a 3 mm thick rectangular aluminum plate. Both plates have the same thickness and approximately the same lateral dimensions. For the determination of the modal shapes, we adopt here a semi-analytical method, the $xyz$ algorithm \citep{migliori1993resonant}. The resulting first 16 eigenmodes and eigenfrequencies calculated using this method are shown in Fig. \ref{fig:modesrectangular}.  

\begin{figure}
    \includegraphics[scale=0.68]{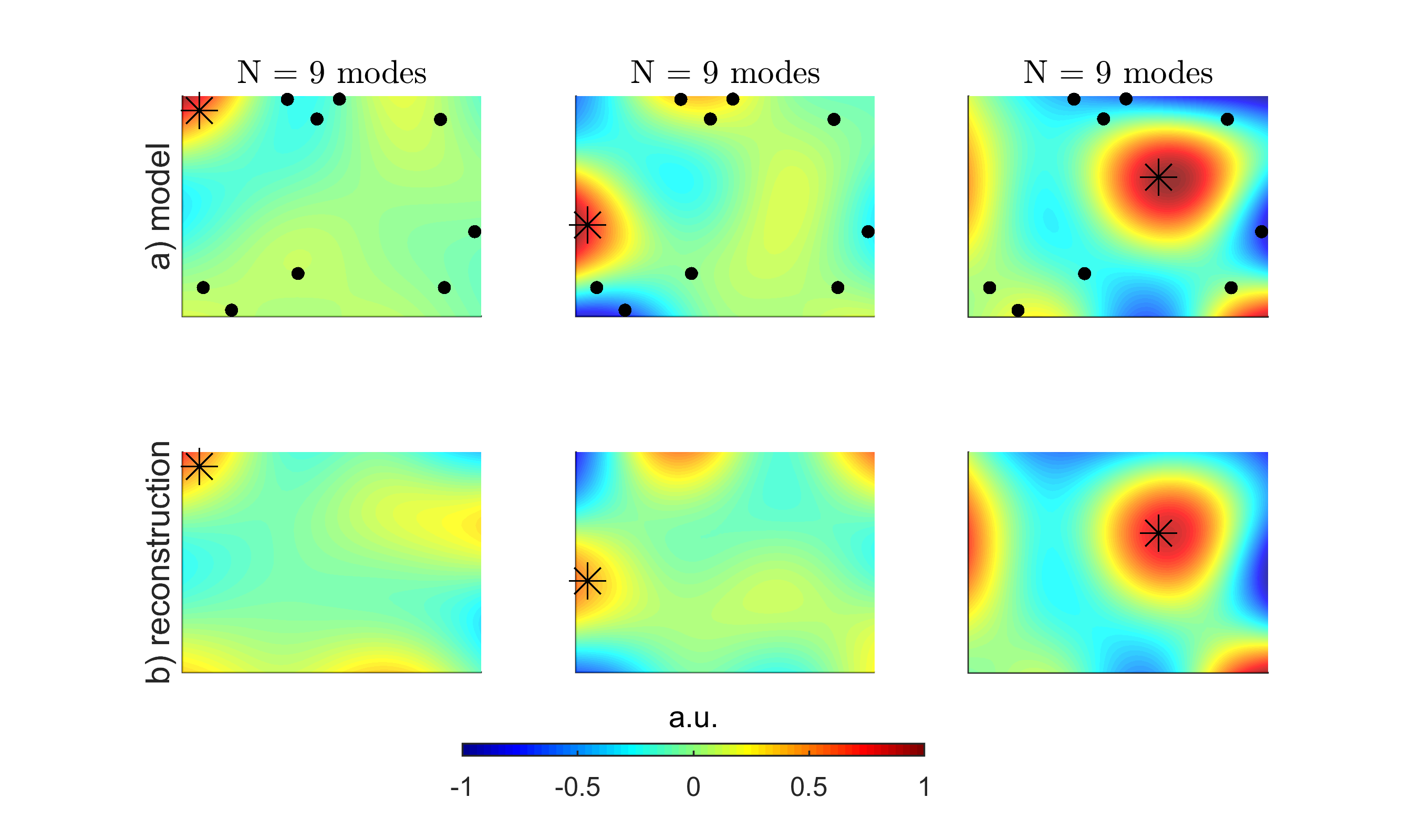}
    \caption{\label{fig:sourcerectangular}Sources localization results in the rectangular plate. a) expected image, built from direct Green's function modal expansion. The black dots represents the used receivers for the imaging, and the title specify the number of modes used for the reconstruction. b) reconstruction with the data. The images are normalized by the maximum of their absolute value.}
\end{figure}

Fig. \ref{fig:sourcerectangular} illustrates graphically the results of source localization in three examples for this plate geometry. There is good agreement between the model and the data. In this experiment, only 9 modes were correctly detected and therefore used in the source reconstruction. The number of points used is also reduced to 9, in order to use equation \ref{eq:reconstruction2}. \\

\section{Monitoring} % around 3.6g and for the square plate around 7.2g

The second experiment presented here illustrates the feasibility of using the modal imaging method for the detection of inhomogeneities in the tested samples. This is realized by placing a small point mass on the surface of the plates and acquiring the same data set as previously (3 sources, 16 receiver points). The test is performed with both the rectangular and the irregularly shaped plates. Experimentally, a simple perturbation of the dynamic response of the plate is obtained by adding a small mass at its surface. The added mass consists of small cubic permanent magnets. The estimated added mass is around 5g for both plates. It is possible to use perturbation theories to model this small change in the structure \citep{agneni1996damage, yam1996theoretical}. As a first approximation, we assume here that a change of mass does not affect the eigenmodes shapes $\left\{\phi_n\right\}$ but only the eigenfrequencies $f_n$ \citep{lott2018three}. This assumption leads to considerable simplifications by neglecting the details of the mechanical interaction of the mass with the plate. Consequently, the mass location can be deduced from the sensitivity of the modes to its addition \citep{fu2001modal}. 

\begin{figure}
    \includegraphics[scale=0.63]{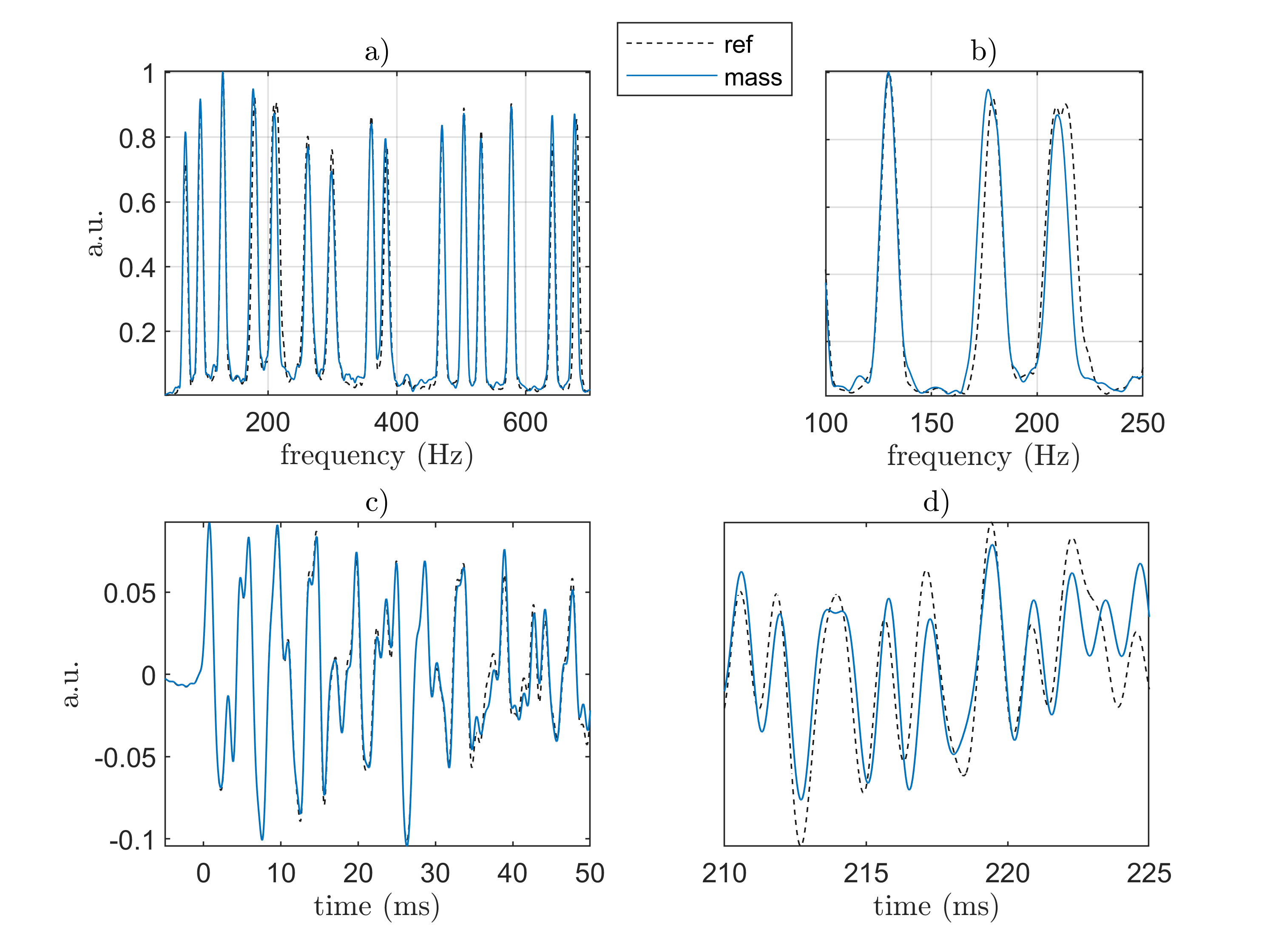}
    \caption{\label{fig:stretch} Signal signatures due to the add of mass. a) The amplitudes and the resonant frequencies are affected by the add of mass. b) (zoom in panel a) The mass decrease the initial resonant frequencies. c) In time domain, the shift of the resonances are translated into a cumulative phase shift in the coda signal. d) same as c) with a zoom in the 30-60 ms time window.}
\end{figure}

We first compare the data set obtained on the pentagonal plate with and without the added mass. The signature of the added mass on the signals is shown in Fig. \ref{fig:stretch}. From the frequency analysis of the data (Fig.\ref{fig:stretch} a-b), the addition of mass induces a shift of the resonant curves to lower frequencies and a modulation of the peak amplitudes. In time domain (Fig.\ref{fig:stretch} c-d), we observe the typical cumulative time shift of the coda\citep{snieder2002coda}, induced here by the changes of the resonant frequencies. From this complex signature, we choose to retain only the frequency shift of the modes as the main localization parameter and propose the following simplified imaging function.

\begin{equation}\label{eq:imagingmassequation}
    I_i = \Sigma_n \dfrac{\delta f_n}{f_n} \left|\transp{\phi_{ni}}\right|
\end{equation}

Where $i$ indicates the pixel of the image to be constructed, $f_n$ the $n^{th}$ resonance frequency for the unperturbed plate, and $\delta f_n$ the variation of a resonance frequency between the reference and perturbed states. 

\begin{figure}
    \includegraphics[scale=0.6]{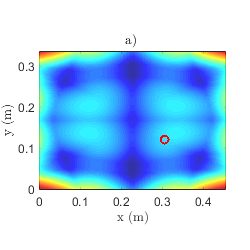}
    \includegraphics[scale=0.6]{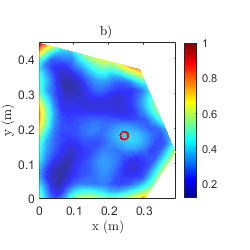}
    \caption{\label{fig:mass} Mass localization results for a) the rectangular plate with an add mass of 7.2g and b) for the pentagonal plate with an add of mass of 3.6g. the mass is a small magnet depicted with the black circle in panels a-b. The images are normalized by the maximum of their absolute value.}
\end{figure}

Results of this imaging procedure are shown in Fig. \ref{fig:mass} for both rectangular (a) and irregular (b) plates. Since perturbation theory is applied on a mode at a time, it is not possible to perform coherent summations over frequencies as in the previous source localization problem. Thus, localizing a change in a structure from the analysis of individual resonances results in symmetry problems during the data post-processing. For a rectangular shaped structure, the incoherent summation over the modes of the wavefield shown in equation \ref{eq:imagingmassequation} leads to four symmetric source images. The irregular pentagonal plate, instead, demonstrates greater uniqueness in the potential location of the mass.

In Fig. \ref{fig:mass}, the edges of the plates show the largest incoherent cumulative amplitudes of vibration. The use of sharp corners for the two plates helps to split the different resonant modes across frequencies, but also induce nonuniform localization of the vibration energy for the resonant modes. Plates with smooth edges provides a more uniform spatial spread of the energy, and thus modal overlapping over frequency \citep{colombi2014green, lott2020localized}, which may add complexity to the modal decomposition technique. The resulting image, built from equation \ref{eq:imagingmassequation}, does not take into account the spatial inhomogeneity of the coherent sum over the modes. This is also a factor preventing access to the mass of the added object, as it should be inverted jointly with its location. \\

\section{Conclusions}

In this paper, we have presented the experimental realization of a source and defect-like elastic imaging procedure, using discrete normal modes of plates with arbitrary geometries. The imaging methodologies are based on a data-driven procedures, which couples numerical simulations with the experimental data. 

First experiment, i.e. the source localization, use modal Green function expansion and composition analytical methods. Sources are shown to be reliably imaged using a coherent sum over the discrete resonant modes supported by the plates, with a mathematical correction of the receivers array geometry. With a limited number of modes and receivers used in the experiment, the resolution of the images are limited but could be applied to any kind of geometry. 

Second imaging problem, i.e. the mass localization, is handle with an incoherent summation over the same modes, and lead to cavity symmetries issues in the resulting images. The incoherent summation of the modes induce four symmetric mass images in the rectangular cavity, and a unique one in the irregular pentagonal cavity. For both cavities, the modal intensity is not homogeneous and create an over-representation of the plates edges in the images. If the vibration amplitude is large enough, the interaction of the mass with the plate can generate harmonics in a nonlinear regime. These harmonics can be used for imaging, by treating the defect as an auxiliary source. However, the defect must demonstrate enough strength in front of the plate in order to be treated as an auxiliary source of nonlinear harmonics. Another important aspect it that the plate is an anharmonic cavity, meaning that normal modes are not necessarily multiples of a fundamental frequencies. The harmonics due to the presence of a defect can alternatively be evanescent, remaining in the defect area, or strongly amplified, if one of the generated harmonics falls on a fundamental mode of the plate. In future, the coherent source imaging technique could be used in reversed mode, using few sources and adapting an array geometry to focus a high level of energy anywhere inside a medium, even at depth, without the need to physically access to the targeted point. We will address this methods in future work.

\section*{Acknowledgements}

All authors are supported by the European Commission under the Future Emerging Technologies Open “Boheme” Grant No. 863179.

\section*{References}

\bibliography{bib}

@article{eiras2021damage,
  title={Damage detection and localization from linear and nonlinear global vibration features in concrete slabs subjected to localized thermal damage},
  author={Eiras, Jes{\'u}s N and Payan, C{\'e}dric and Rakotonarivo, Sandrine and Garnier, Vincent},
  journal={Structural Health Monitoring},
  volume={20},
  number={2},
  pages={567--579},
  year={2021},
  publisher={SAGE Publications Sage UK: London, England}
}

@article{van2007multi,
  title={Multi-mode nonlinear resonance ultrasound spectroscopy for defect imaging: An analytical approach for the one-dimensional case},
  author={Van Den Abeele, Koen},
  journal={The Journal of the Acoustical Society of America},
  volume={122},
  number={1},
  pages={73--90},
  year={2007},
  publisher={Acoustical Society of America}
}

@article{cawley1979location,
  title={The location of defects in structures from measurements of natural frequencies},
  author={Cawley, Peter and Adams, Robert Darius},
  journal={The Journal of Strain Analysis for Engineering Design},
  volume={14},
  number={2},
  pages={49--57},
  year={1979},
  publisher={SAGE Publications Sage UK: London, England}
}

@article{farrar1998comparative,
  title={Comparative study of damage identification algorithms applied to a bridge: I. Experiment},
  author={Farrar, Charles R and Jauregui, David A},
  journal={Smart materials and structures},
  volume={7},
  number={5},
  pages={704},
  year={1998},
  publisher={IOP Publishing}
}

@article{humar2006performance,
  title={Performance of vibration-based techniques for the identification of structural damage},
  author={Humar, Jag and Bagchi, Ashutosh and Xu, Hongpo},
  journal={Structural Health Monitoring},
  volume={5},
  number={3},
  pages={215--241},
  year={2006},
  publisher={Sage Publications Sage CA: Thousand Oaks, CA}
}

@article{roux2014structural,
  title={Structural-change localization and monitoring through a perturbation-based inverse problem},
  author={Roux, Philippe and Gu{\'e}guen, Philippe and Baillet, Laurent and Hamze, Alaa},
  journal={The Journal of the Acoustical Society of America},
  volume={136},
  number={5},
  pages={2586--2597},
  year={2014},
  publisher={Acoustical Society of America}
}

@article{gallot2011coherent,
  title={Coherent backscattering enhancement in cavities. Highlights of the role of symmetry},
  author={Gallot, Thomas and Catheline, Stefan and Roux, Philippe},
  journal={The Journal of the Acoustical Society of America},
  volume={129},
  number={4},
  pages={1963--1971},
  year={2011},
  publisher={Acoustical Society of America}
}

@article{prada1996decomposition,
  title={Decomposition of the time reversal operator: Detection and selective focusing on two scatterers},
  author={Prada, Claire and Manneville, S{\'e}bastien and Spoliansky, Dimitri and Fink, Mathias},
  journal={The Journal of the Acoustical Society of America},
  volume={99},
  number={4},
  pages={2067--2076},
  year={1996},
  publisher={Acoustical Society of America}
}

@article{fink2000time,
  title={Time-reversed acoustics},
  author={Fink, Mathias and Cassereau, Didier and Derode, Arnaud and Prada, Claire and Roux, Philippe and Tanter, Mickael and Thomas, Jean-Louis and Wu, Fran{\c{c}}ois},
  journal={Reports on progress in Physics},
  volume={63},
  number={12},
  pages={1933},
  year={2000},
  publisher={IOP Publishing}
}

@article{migliori1993resonant,
  title={Resonant ultrasound spectroscopic techniques for measurement of the elastic moduli of solids},
  author={Migliori, A\_ and Sarrao, JL and Visscher, William M and Bell, TM and Lei, Ming and Fisk, Z and Leisure, R Gi},
  journal={Physica B: Condensed Matter},
  volume={183},
  number={1-2},
  pages={1--24},
  year={1993},
  publisher={Elsevier}
}

@article{remillieux2015resonant,
  title={Resonant ultrasound spectroscopy for materials with high damping and samples of arbitrary geometry},
  author={Remillieux, Marcel C and Ulrich, TJ and Payan, C{\'e}dric and Rivi{\`e}re, Jacques and Lake, Colton R and Le Bas, Pierre-Yves},
  journal={Journal of Geophysical Research: Solid Earth},
  volume={120},
  number={7},
  pages={4898--4916},
  year={2015},
  publisher={Wiley Online Library}
}

@article{zadler2004resonant,
  title={Resonant ultrasound spectroscopy: theory and application},
  author={Zadler, Brian J and Le Rousseau, J{\'e}r{\^o}me HL and Scales, John A and Smith, Martin L},
  journal={Geophysical Journal International},
  volume={156},
  number={1},
  pages={154--169},
  year={2004},
  publisher={Blackwell Publishing Ltd Oxford, UK}
}

@article{mizusawa1986natural,
  title={Natural frequencies of rectangular plates with free edges},
  author={Mizusawa, T},
  journal={Journal of Sound and Vibration},
  volume={105},
  number={3},
  pages={451--459},
  year={1986},
  publisher={Elsevier}
}

@article{wang2004static,
  title={Static and free vibration analyses of rectangular plates by the new version of the differential quadrature element method},
  author={Wang, Yongliang and Wang, Xinwei and Zhou, Yong},
  journal={International Journal for Numerical Methods in Engineering},
  volume={59},
  number={9},
  pages={1207--1226},
  year={2004},
  publisher={Wiley Online Library}
}

@article{fasi2015algorithm,
  title={An algorithm for the matrix Lambert W function},
  author={Fasi, Massimiliano and Higham, Nicholas J and Iannazzo, Bruno},
  journal={SIAM Journal on Matrix Analysis and Applications},
  volume={36},
  number={2},
  pages={669--685},
  year={2015},
  publisher={SIAM}
}

@article{touma2021distortion,
  title={A distortion matrix framework for high-resolution passive seismic 3-D imaging: application to the San Jacinto fault zone, California},
  author={Touma, Rita and Blondel, Thibaud and Derode, Arnaud and Campillo, Michel and Aubry, Alexandre},
  journal={Geophysical Journal International},
  volume={226},
  number={2},
  pages={780--794},
  year={2021},
  publisher={Oxford University Press}
}

@article{astorga2019recovery,
  title={Recovery of the resonance frequency of buildings following strong seismic deformation as a proxy for structural health},
  author={Astorga, Ariana Lucia and Gueguen, Philippe and Riviere, Jacques and Kashima, Toshihide and Johnson, Paul Allan},
  journal={Structural Health Monitoring},
  volume={18},
  number={5-6},
  pages={1966--1981},
  year={2019},
  publisher={Sage Publications Sage UK: London, England}
}

@article{senjanovic2016analytical,
  title={An analytical solution to free rectangular plate natural vibrations by beam modes--ordinary and missing plate modes},
  author={Senjanovi{\'c}, Ivo and Tomi{\'c}, Marko and Vladimir, Nikola and Had{\v{z}}i{\'c}, Neven},
  journal={Transactions of FAMENA},
  volume={40},
  number={3},
  pages={1--18},
  year={2016},
  publisher={Fakultet strojarstva i brodogradnje}
}

@inproceedings{agneni1996damage,
  title={Damage Detection on Aeronautical Structures by a Mixed Approach in the Frequency Domain},
  author={Agneni, A and Balis Crema, L and Castellani, A and Mastroddi, F},
  journal={Proceedings of the 14th International Modal Analysis Conference},
  volume={2768},
  pages={1415},
  year={1996}
}

@article{robert2008green,
  title={Green’s function estimation in speckle using the decomposition of the time reversal operator: Application to aberration correction in medical imaging},
  author={Robert, Jean-Luc and Fink, Mathias},
  journal={The Journal of the Acoustical Society of America},
  volume={123},
  number={2},
  pages={866--877},
  year={2008},
  publisher={Acoustical Society of America}
}

@article{aubry2009detection,
  title={Detection and imaging in a random medium: A matrix method to overcome multiple scattering and aberration},
  author={Aubry, Alexandre and Derode, Arnaud},
  journal={Journal of Applied Physics},
  volume={106},
  number={4},
  pages={044903},
  year={2009},
  publisher={American Institute of Physics}
}

@article{aubry2014recurrent,
  title={Recurrent scattering and memory effect at the Anderson localization transition},
  author={Aubry, Alexandre and Cobus, Laura A and Skipetrov, Sergey E and Van Tiggelen, Bart A and Derode, Arnaud and Page, John H},
  journal={Physical review letters},
  volume={112},
  number={4},
  pages={043903},
  year={2014},
  publisher={APS}
}

@article{seydoux2017pre,
  title={Pre-processing ambient noise cross-correlations with equalizing the covariance matrix eigenspectrum},
  author={Seydoux, L{\'e}onard and de Rosny, Julien and Shapiro, Nikolai M},
  journal={Geophysical Journal International},
  volume={210},
  number={3},
  pages={1432--1449},
  year={2017},
  publisher={Oxford University Press}
}

@article{poggi2010estimating,
  title={Estimating Rayleigh wave particle motion from three-component array analysis of ambient vibrations},
  author={Poggi, Valerio and F{\"a}h, Donat},
  journal={Geophysical Journal International},
  volume={180},
  number={1},
  pages={251--267},
  year={2010},
  publisher={Blackwell Publishing Ltd Oxford, UK}
}

@article{lambert2020distortion,
  title={Distortion matrix approach for ultrasound imaging of random scattering media},
  author={Lambert, William and Cobus, Laura A and Frappart, Thomas and Fink, Mathias and Aubry, Alexandre},
  journal={Proceedings of the National Academy of Sciences},
  volume={117},
  number={26},
  pages={14645--14656},
  year={2020},
  publisher={National Acad Sciences}
}

@article{yam1996theoretical,
  title={Theoretical and experimental study of modal strain analysis},
  author={Yam, LY and Leung, TP and Li, DB and Xue, KZ},
  journal={Journal of Sound and vibration},
  volume={191},
  number={2},
  pages={251--260},
  year={1996},
  publisher={Elsevier}
}

@article{lott2018three,
  title={Three-dimensional modeling and numerical predictions of multimodal nonlinear behavior in damaged concrete blocks},
  author={Lott, Martin and Payan, Cedric and Garnier, Vincent and Le Bas, Pierre-Yves and Ulrich, Timothy James and Remillieux, Marcel C},
  journal={The Journal of the Acoustical Society of America},
  volume={144},
  number={3},
  pages={1154--1159},
  year={2018},
  publisher={Acoustical Society of America}
}

@article{baggeroer1988matched,
  title={Matched field processing: Source localization in correlated noise as an optimum parameter estimation problem},
  author={Baggeroer, Arthur B and Kuperman, WA and Schmidt, Henrik},
  journal={The Journal of the Acoustical Society of America},
  volume={83},
  number={2},
  pages={571--587},
  year={1988},
  publisher={Acoustical Society of America}
}

@article{collins1991focalization,
  title={Focalization: Environmental focusing and source localization},
  author={Collins, Michael D and Kuperman, WA},
  journal={The Journal of the Acoustical Society of America},
  volume={90},
  number={3},
  pages={1410--1422},
  year={1991},
  publisher={Acoustical Society of America}
}

@book{fu2001modal,
  journal={Modal analysis, chapter 11.3},
  author={Fu, Zhi-Fang and He, Jimin},
  year={2001},
  publisher={Elsevier}
}

@article{colombi2014temporal,
  title={On the temporal stability of the coda of ambient noise correlations},
  author={Colombi, Andrea and Chaput, Julien and Brenguier, Florent and Hillers, Gregor and Roux, Philippe and Campillo, Michel},
  journal={Comptes Rendus Geoscience},
  volume={346},
  number={11-12},
  pages={307--316},
  year={2014},
  publisher={Elsevier}
}

@article{snieder2002coda,
  title={Coda wave interferometry for estimating nonlinear behavior in seismic velocity},
  author={Snieder, Roel and Gr{\^e}t, Alexandre and Douma, Huub and Scales, John},
  journal={Science},
  volume={295},
  number={5563},
  pages={2253--2255},
  year={2002},
  publisher={American Association for the Advancement of Science}
}

@article{lott2020localized,
  title={Localized modes on a metasurface through multiwave interactions},
  author={Lott, Martin and Roux, Philippe and Seydoux, L{\'e}onard and Tallon, Benoit and Pelat, Adrien and Skipetrov, Sergey and Colombi, Andrea},
  journal={Physical Review Materials},
  volume={4},
  number={6},
  pages={065203},
  year={2020},
  publisher={APS}
}

@article{colombi2014green,
  title={Green's function retrieval through cross-correlations in a two-dimensional complex reverberating medium},
  author={Colombi, Andrea and Boschi, Lapo and Roux, Philippe and Campillo, Michel},
  journal={The Journal of the Acoustical Society of America},
  volume={135},
  number={3},
  pages={1034--1043},
  year={2014},
  publisher={Acoustical Society of America}
}

@article{poli2017analysis,
  title={Analysis of intermediate period correlations of coda from deep earthquakes},
  author={Poli, Piero and Campillo, Michel and de Hoop, Maarten},
  journal={Earth and Planetary Science Letters},
  volume={477},
  pages={147--155},
  year={2017},
  publisher={Elsevier}
}

@article{larose2010locating,
  title={Locating a small change in a multiple scattering environment},
  author={Larose, Eric and Planes, Thomas and Rossetto, Vincent and Margerin, Ludovic},
  journal={Applied Physics Letters},
  volume={96},
  number={20},
  pages={204101},
  year={2010},
  publisher={American Institute of Physics}
}

@article{catheline2011coherent,
  title={Coherent backscattering enhancement in cavities: The simple-shape cavity revisited},
  author={Catheline, Stefan and Gallot, Thomas and Roux, Philippe and Ribay, Guillemette and De Rosny, Julien},
  journal={Wave motion},
  volume={48},
  number={3},
  pages={214--222},
  year={2011},
  publisher={Elsevier}
}

@article{weaver1994weak,
  title={Weak Anderson localization and enhanced backscatter in reverberation rooms and quantum dots},
  author={Weaver, Richard L and Burkhardt, John},
  journal={The Journal of the Acoustical Society of America},
  volume={96},
  number={5},
  pages={3186--3190},
  year={1994},
  publisher={Acoustical Society of America}
}

@article{gliozzi2006modelling,
  title={Modelling localized nonlinear damage and analysis of its influence on resonance frequencies},
  author={Gliozzi, AS and Nobili, M and Scalerandi, M},
  journal={Journal of Physics D: Applied Physics},
  volume={39},
  number={17},
  pages={3895},
  year={2006},
  publisher={IOP Publishing}
}

\end{document}